\documentclass[aps,prb,reprint,groupedaddress,twocolumn,showkeys,amsmath,amssymb,longbibliography,floatfix]{revtex4-1}
\usepackage[english]{babel}
\usepackage{amsmath,amssymb,mathrsfs,bm,braket,color,graphicx,comment,amsfonts,mwe,tikz}
\usepackage{tabularx}
\usepackage[export]{adjustbox}
\usepackage[percent]{overpic}

\usepackage{blindtext}
\usepackage{diagbox}
\usepackage{todonotes}
\usepackage{contour,color,stackengine,titletoc,lipsum,tikz}
\usepackage[colorlinks,citecolor=blue,urlcolor=blue]{hyperref}
\usepackage{float}
\usepackage[mathscr]{euscript}	
\usepackage[normalem]{ulem}

\newcommand{\vect}[1]{\boldsymbol{#1}}

\newcommand{\mf}[1]{\mathbf{#1}}

\begin{document}
\title{Entanglement spectroscopy of anomalous surface states}

\author{Arjun Dey}
\affiliation{Department of Condensed Matter Physics, Weizmann Institute of Science, Rehovot 7610001, Israel}

\author{David F. Mross}
\affiliation{Department of Condensed Matter Physics, Weizmann Institute of Science, Rehovot 7610001, Israel}

\begin{abstract}
We study entanglement spectra of gapped states on the surfaces of symmetry-protected topological phases. These surface states carry anomalies that do not allow them to be terminated by a trivial state. Their entanglement spectra are dominated by non-universal features, which reflect the underlying bulk. We introduce a modified type of entanglement spectra that incorporate the anomaly and argue that they correspond to physical edge states between different surface states. We support these arguments by explicit analytical and numerical calculations for free and interacting surfaces of three-dimensional topological insulators of electrons.

\end{abstract}

\maketitle

{\bf Introduction}. Over the past years, entanglement has surpassed the notion of correlations for characterizing and identifying quantum many-body systems. Famously, the entanglement entropy of two-dimensional gapped systems contains a universal contribution that is non-trivial for topologically ordered states\cite{Preskill+Kitaev2006,Levin+Wen2006}. This contribution vanishes for symmetry-protected topological states (SPTs), which do not host any fractional quasiparticles in the bulk. Still, their non-trivial topological nature can be deduced by resolving the entanglement entropy according to symmetries\cite{symresentgoldstein2018,symresentxavier2018,symresentbonsignori2019}.

More refined information about topological states can be obtained from entanglement spectra (ES). In a seminal work, Li and Haldane \cite{HaldaneES} showed that the ES of certain quantum Hall states and their energy spectra at a physical edge are describable by the same conformal field theory. They argued that topological phases can thus be identified by their ES. Subsequently, such an ES--edge state correspondence has been proven for a broad class of two-dimensional topological states\cite{ESedgeLudwig2012, BulkEdgeESBernevig2011}. Additionally, the agreement of ES and physical edge spectra has been confirmed empirically for various other systems, including topological insulators\cite{TIViswanath2010, TIFidkowski2010}, $p$-wave superfluids in the continuum\cite{ChiralSuperfluidsReadDubail} and on a lattice\cite{LatticePwaveES2009, Kim_2014Pwavees}, fractional quantum Hall states\cite{FQHEJainHaldane2009, FQHETorus2010, FQHEThomaleBernevig2010}, spin chains\cite{SpinChainThomaleBernevig2010} and the Kitaev honeycomb model\cite{KitaevChainES2010}.

\begin{figure}[ht]
 \includegraphics[width=\columnwidth]{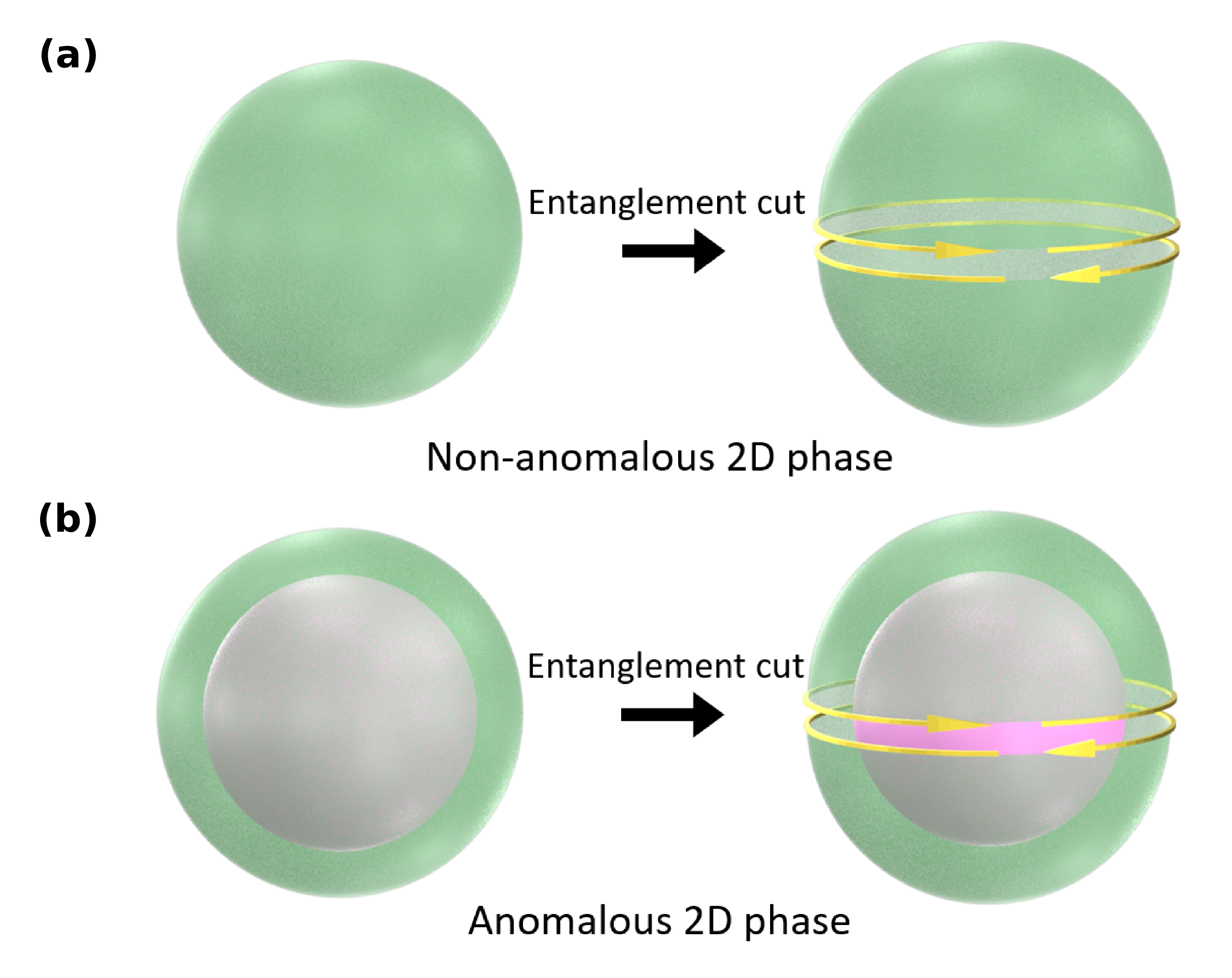}
 % \vspace{-2mm}
 \caption{In non-anomalous systems, a real-space entanglement cut corresponds to an analogous physical cut, see panel (a). Anomalous states are inextricably tied to a topologically non-trivial bulk, which is exposed upon performing a cut, see panel (b).}
 \vspace{-4mm}
 \label{fig: anomalous and non-anomalous}
\end{figure}

The conjectured ES--edge correspondence makes the tacit assumption that a physical edge is possible. As such, it does not directly apply to an essential and widely studied class of condensed-matter systems:
Surface states of topological insulators or superconductors. Such states cannot be `stripped' from their host. Physically removing a finite surface layer of any topological system exposes its bulk, leading to the formation of a new surface state. We expect that, similarly, real-space entanglement cuts of such systems reveal the underlying bulk, and the ES are dominated by bulk properties (cf.~Fig.~\ref{fig: anomalous and non-anomalous}). 

More formally, any state that can arise on the surface of a given SPT carries the same anomaly. If the SPT is non-trivial, this anomaly is incompatible with a trivial vacuum state, and the surface state cannot be terminated by a physical edge. Only boundaries between surface states with the same anomaly are possible. Crucially, the interface between two gapped surface phases hosts topologically protected states, uniquely identifying one provided the other is known. In this paper, we show how entanglement spectroscopy can determine these edge states between an unknown state and an arbitrary free-fermion state.

As a concrete system for our numerical calculations, we use an electronic topological insulator (TI) with time-reversal symmetry ${\cal T}^2=-1$\cite{Hasan3DTI_2010, Hasan3DTI_2011, Kane3DTI_2011} as the paradigmatic example of a 3D SPT. When its two-dimensional surface is symmetric and non-interacting, it hosts a single two-component Dirac fermion\cite{DiracCone1, DiracCone2}. This theory cannot arise in strictly two-dimensional systems with the same symmetries due to fermion doubling~\cite{FermionDoubling1, FermionDoubling2, FermionDoubling3}. When time-reversal symmetry is broken on the surface, the Dirac fermions become massive and realize a surface Hall conductance of $\sigma_{xy}=\frac{1}{2}$\cite{TRSbreaking1, TRSbreaking2} (in units of $\frac{e^2}{h}$). By contrast, the Hall conductance of gapped free fermion systems in strictly two-dimensions must be an integer. Similarly, breaking charge conservation realizes a time-reversal-invariant cousin of topological $p+ip$\cite{Pwave3, Pwave1, Pwave2} superconductors with Majorana modes in vortex cores. Finally, strong interactions may gap the surface while preserving both symmetries by forming an anomalous topological order\cite{StrongInteractions1, StrongInteractions2, StrongInteractions3, StrongInteractions4, StrongInteractions5}.

{\bf Entanglement spectra of anomalous surfaces}. Any surface wave function of a free-fermion SPT may be expressed as 
\begin{equation}
 |\Phi\rangle_\Lambda =\hat\Phi_\Lambda(c^\dagger_{E,i}) |0\rangle~.\label{surfacewf}
\end{equation}
Here $\hat\Phi_\Lambda$ is an operator-valued function, and $c^\dagger_{E,i}$ creates an electron in a single-particle eigenstate with energy $E$, and additional quantum numbers $i$. The `empty state' $|0\rangle$ denotes a Fermi sea filled up to the top of the valence band at energy $\Lambda$. The symmetric, non-interacting surface is thus represented by $ | \Phi \rangle_\Lambda^0 =\prod_i \prod_{E=\Lambda}^\mu c^\dagger_{E,i} \ket{0} $ with chemical potential $\mu$. In general, $ \ket{\Phi}_{\Lambda} $ must reduce to $ \ket{\Phi}_{\Lambda}^0 $ for $c_{E \rightarrow \Lambda}$ to describe a surface state that does not hybridize with the bulk. The cutoff $\Lambda$ is inevitable due to the anomalous nature of the surface state, but its precise value does not affect local observables.

\begin{figure}
 \includegraphics[width=\columnwidth]{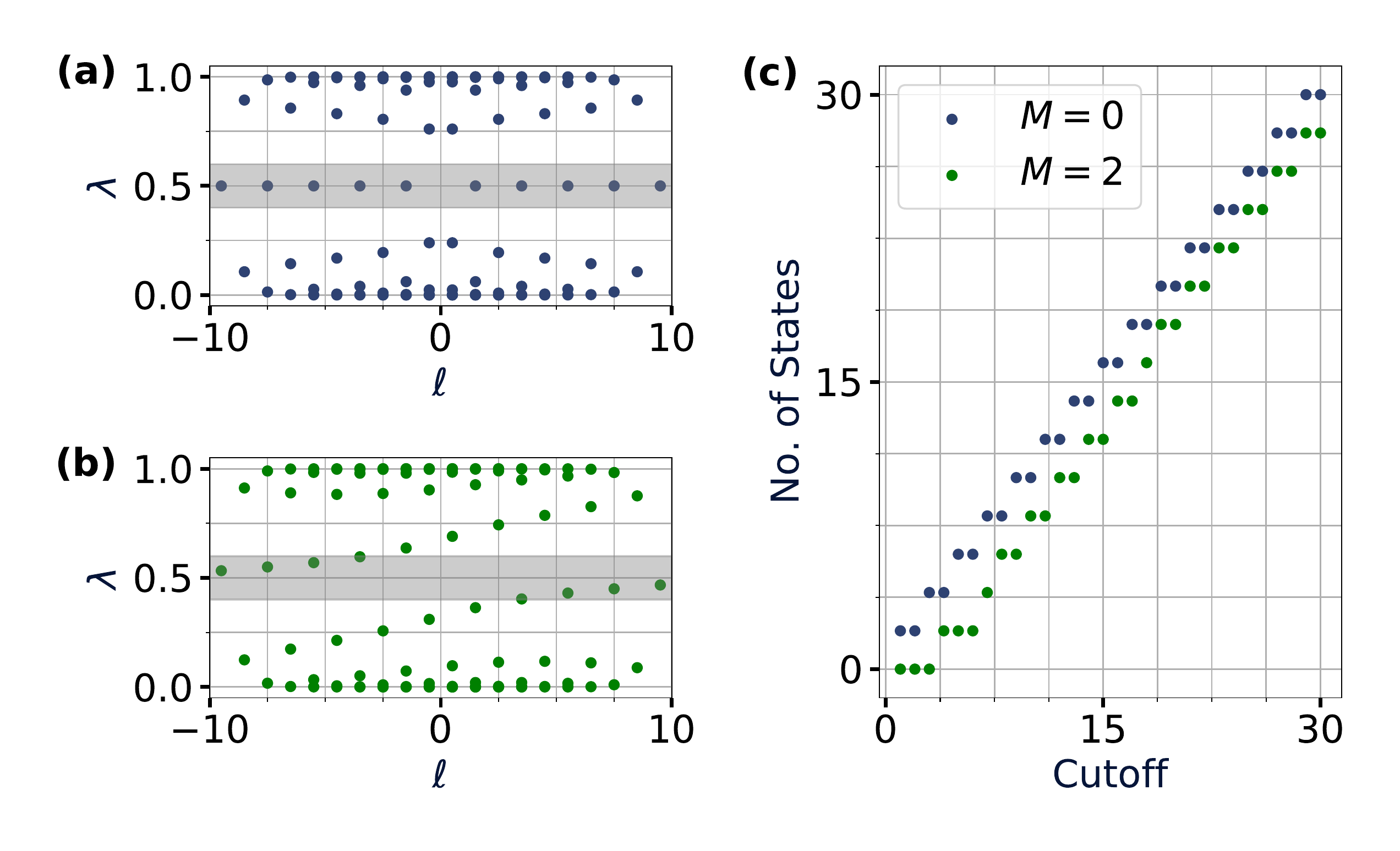}
 \caption{The entanglement spectra of anomalous surface states exhibit a large number of low-lying states ($\lambda_i \approx 0.5$), that increases with the cutoff. Panels (a) and (b) show the spectra for a gapless and massive Dirac cone, respectively, with cutoff $\Lambda=10$. In both cases, the number of states within the shaded low-energy window scales linearly with the cutoff and is thus non-universal (c).}
 \vspace{-4mm}
 \label{fig: cutoff dependence of vanilla ES}
\end{figure}

To test our expectations for the ES of anomalous states, we study the surface of a 3D TI, which hosts a single Dirac cone governed by
\begin{equation}
 H_{M} =
 \sum _{\vect k}
\phi_{\vect k}^
 \dagger [\vect k \cdot \vect \sigma + M \sigma_z]\phi_{\vect k}
 ~.\label{eqn.massivedirac}
\end{equation}
Here, $\phi_{\vect k}^\dagger$ creates electrons with momentum $\vect{k}=(k_x,k_y)$, $\sigma_x,\sigma_y,\sigma_z$ are Pauli matrices, and $M$ is a time-reversal symmetry-breaking mass. The un-normalized single-particle eigenstates are
\begin{equation}
 \vect{v}_{\vect k}= \begin{pmatrix}
\sqrt{{\epsilon+M}}\\\sqrt{{\epsilon-M}}e^{-i \varphi_{\mathbf{k}}}
 \end{pmatrix},\quad
\vect{u}_{\vect k} =
 \begin{pmatrix}
\sqrt{{\epsilon-M}}e^{i \varphi_{\mathbf{k}}}\\
- \sqrt{\epsilon+M}
 \end{pmatrix},\label{eqn.uandv}
\end{equation}
where $\epsilon = \sqrt{k^2+M^2}$ and $k,\varphi_{\vect k}$ are the magnitude and polar angle of the two-dimensional momentum. The corresponding single-particle energies are $E_{\vect v}=-E_{\vect u}=\epsilon$, and the ground state of $H_{M}$ is $|\Phi^M\rangle_{\Lambda} = \prod_{\vect k}\vect u_{\vect k} \phi_{\vect{k}}^{\dagger} \ket{0} $. In Appendix~\ref{app: surface states of spherical 3DTI spectrum}, we show the spectrum of Eq.~(\ref{eqn.massivedirac}) for spherical geometry.

We obtain the ES by decomposing the Hilbert space of the surface
into two parts, ${\cal H}={\cal H}_\alpha \otimes {\cal H}_\beta$. The Schmidt decomposition
\begin{equation}
|\Phi\rangle =\sum\nolimits_n e^{-\lambda_n/2} |\Phi_{\alpha,n}\rangle\otimes |\Phi_{\beta,n}\rangle~,
\end{equation}
with $|\Phi_{\alpha(\beta),n}\rangle\in {\cal H}_{\alpha(\beta)}$ yields the `entanglement energies' $\lambda_i$. We numerically compute these numbers for $|\Phi^M\rangle_{\Lambda}$ with $\alpha,\beta$, the two hemispheres of a sphere with radius $R$ and (Fig.~\ref{fig: cutoff dependence of vanilla ES}). For local observables, we would expect universal results when $R^{-1}\ll M \ll \Lambda$. For the ES, we instead find that the number of low-lying $\lambda_i$ grows linearly with the cutoff $\Lambda$. In Fig.~\ref{fig: cutoff dependence of vanilla ES} panel (c), we show the number of pseudo energy states as a function of the cutoff for massless and massive Dirac fermions. We attribute this cutoff dependence and the large number of low-lying states to the exposure of the underlying bulk (cf.~Fig.~\ref{fig: anomalous and non-anomalous}). As anticipated in the introduction, straightforward computation of a surface state's ES is not suitable for its identification.

{\bf Relative entanglement spectra}. 
We adapt entanglement spectroscopy to SPT surfaces by drawing on insights into physical surface spectra (cf. Appendix \ref{app: physical energy spectrum of spherical 3DTI}). Recall that edge-energy spectra are only meaningful for two surface states with the same anomaly. However, their boundary state can equivalently arise at the physical edge of a specific non-anomalous phase, which is subject to the standard ES-edge correspondence.
Our strategy is thus to construct non-anomalous wave functions that encode the boundary between two surface states $A$ and $B$. 

To obtain the desired wave functions, we begin with a gapped free-fermion surface state $A$, whose particle- and hole-like excitations are created by $c^{A,\dagger}_{+,i},c^{A,\dagger}_{-,i}$. By construction, $A$ corresponds to a non-anomalous wave function of particles and holes, i.e., their trivial vacuum. Next, we consider a puddle of any other surface state embedded within $A$. It can be created locally from particle and hole excitations on top of the uniform $A$ state, which does not introduce any anomaly. Likewise, any surface state $B$ can be encoded in a non-anomalous wave function of the excitations $c^A_{\pm,i}$. Such wave functions depend on both $A$ and $B$; we thus refer to them as \textit{relative wave functions} from which we obtain the \textit{relative ES}. We now proceed by numerically calculating the relative ES of various 3D TI surface states and comparing them with the physical edge spectra. Subsequently, we elaborate on the relative wave functions and provide an analytical perspective on their ES.

{\bf Numerical results I: Free fermions.} We compute the relative ES described above for various surface states of a spherical 3D TI. For $|\Phi^A \rangle_{\Lambda}$, we take states with a magnetic or superconducting gap, i.e., the ground states of Eq.~\eqref{eqn.massivedirac} or of 
\begin{equation}
H^\text{SC}_{\Delta_s} =H_0+\Delta_s \sum\limits_{\vect k}\left[\phi_{\vect k,\uparrow}\phi_{-\vect k,\downarrow} + \text{H.c.}\right]~.
\label{eqn.scprox}
\end{equation}
To establish that the relative ES reproduces the boundary between anomalous surfaces and is cutoff independent, we also take the second state to be one of free fermions. In this case, the ES can be efficiently computed using the correlation matrix of particles and holes \cite{correlationMatrix, correlationMatrix2, correlationMatrix3, correlationMatrix4, correlationMatrix5, correlationMatrixBdG, correlationMatrixBdG2, correlationMatrixBdG3} (see also Appendix~\ref{app: zeroes of the decomposed hamiltonian}). The corresponding physical energy spectra are well known. They are summarized, e.g., in Sec.~V of the review Ref.~\onlinecite{Hasan3DTI_2010} and shown for the spherical geometry in Appendix~\ref{app: physical energy spectrum of spherical 3DTI}.

\begin{figure}[h]
 \includegraphics[width=\columnwidth]{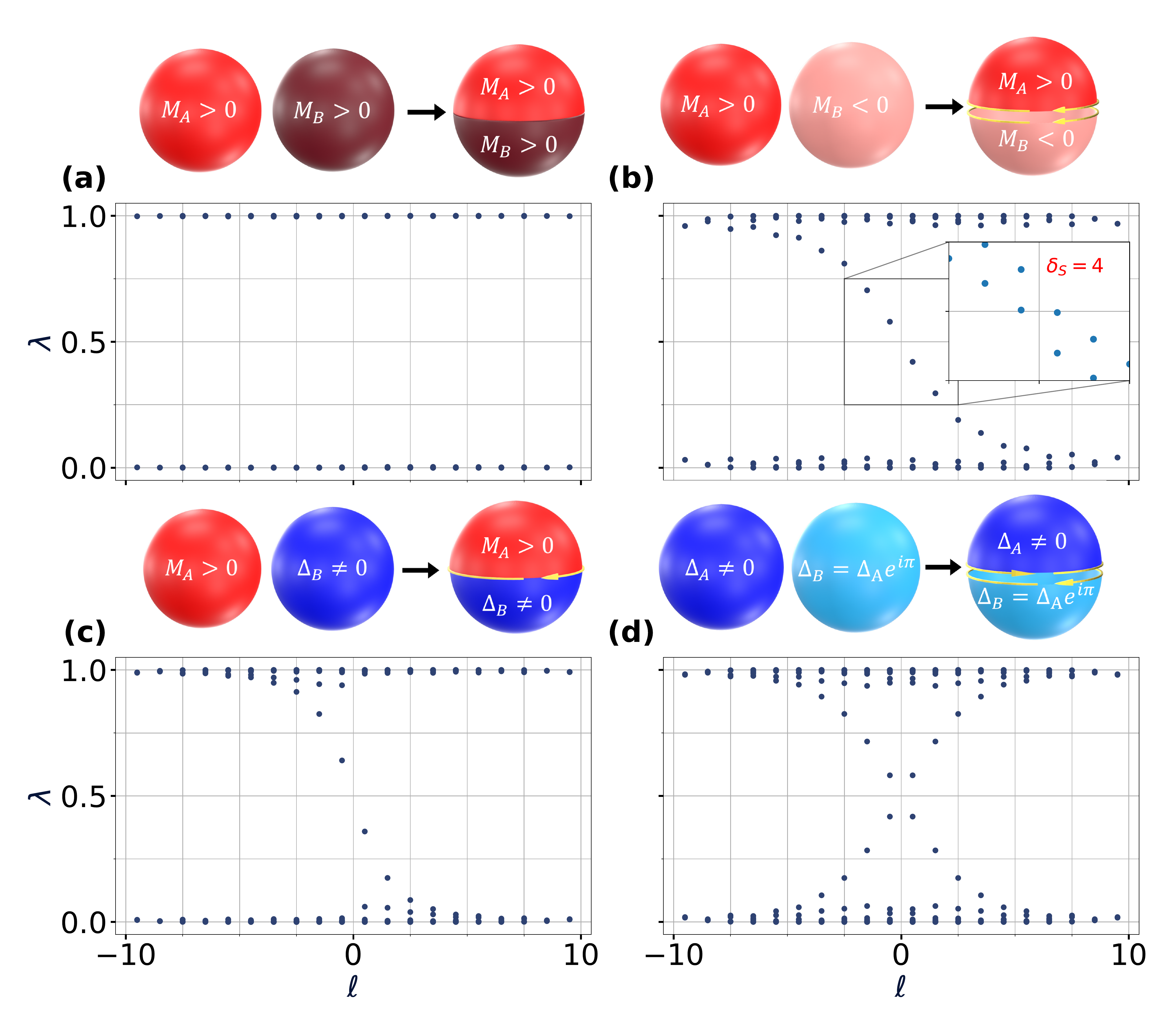}
 \vspace{-4mm}
 \caption{The relative entanglement spectra of different 3D topological insulator surface states match the energy spectra at a physical boundary between the same states. For magnetically gapped states with the same sign, the entanglement spectrum is gapped, as shown in (a) for $M_A=4$ and $M_B=6$. For opposite signs, there is a chiral edge state, shown in (b) for $M_A=4$ and $M_B=-4$. The edge state corresponds to a complex fermion and is split by breaking charge conservation (inset). Similarly, the relative entanglement spectrum of a magnetically gapped and a paired state exhibits a single chiral Majorana mode, shown in (c) for $M_A=4,\Delta_B=-4$. For two paired states with $\pi$ phase difference we find a non-chiral state with helical Majorana fermions, see (d) where $\Delta_A=2,\Delta_B=-8$. 
 }
 \label{fig: reles for non-interacting systems}
\end{figure}

Fig.~\ref{fig: reles for non-interacting systems} shows the relative ES for various choices of the gapped free-fermion systems $A,B$. 
In Fig.~\ref{fig: reles for non-interacting systems}(a), we take $A$ and $B$ to be different representatives of the same phase. Their relative ES is gapped, as expected. In (b), we take both $A$ and $B$ as magnetically gapped, but with opposite signs. Here, the ES describes a chiral Dirac fermion, matching a physical boundary between the same phases. Panel (c) depicts the relative ES between a state with a magnetic gap and a second with a superconducting gap. The Majorana mode in the ES matches the physical energy spectrum at such a boundary. Finally, panel (d) shows the case where $A$ and $B$ are superconductors with a phase difference $\pi$. Again, the ES correctly reproduces the expected helical Majorana edge states. Additional data about the cutoff dependence of these spectra and their dependence on the magnitude are shown in Appendices~\ref{app: cutoff dependence of es and reles} and \ref{app: evolution of reles}. In Appendix~\ref{app: ring geometry 2D qsh}, we show the relative ES for a $1$D anomalous edge state.

\begin{figure}
 % \vspace{2mm}
 \includegraphics[width=\columnwidth]{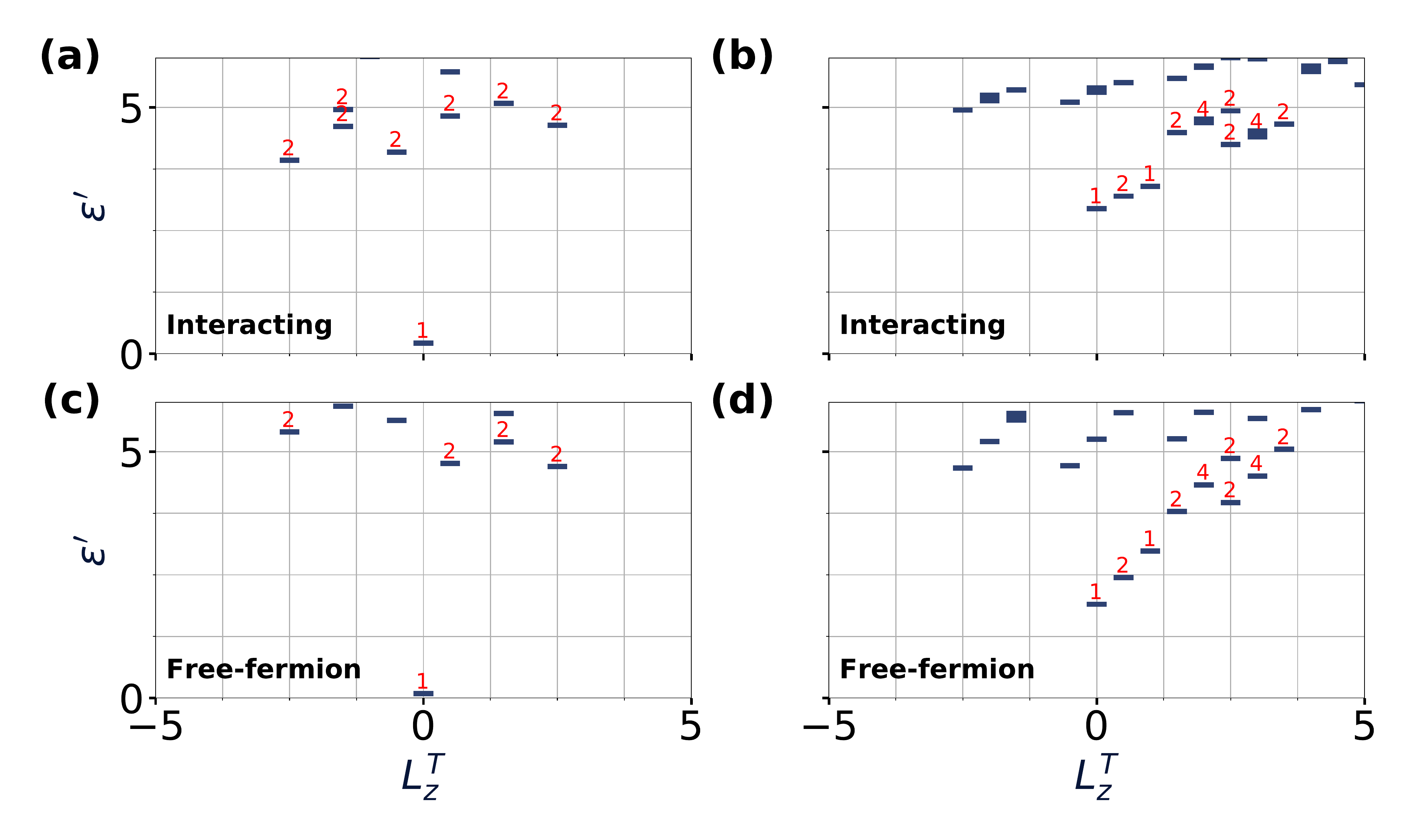}
 \vspace{-6mm}
 \caption{The relative entanglement spectra between repulsive Dirac electrons and massive non-interacting electrons clearly identify the phase of the former. When the sign of the spontaneous magnetization matches the mass of the free-fermion state, the entanglement spectrum is gapped (a). When they are opposite, there is a single chiral state (b). Panels (c) and (d) show the corresponding many-body entanglement spectra for free-fermion wave functions with the same cutoff. The degeneracies of the pseudo energies are indicated in red.}
 \label{fig: Many-body RelES cutoff3}
\end{figure}

{\bf Numerical results II: Interacting states.}
We verify that the relative ES extends beyond free-fermion states by studying Dirac electrons with contact interactions $U$, i.e., the model
\begin{align}
H_\text{Int} = H_0 + U\int_{\vect r \in S^2} \phi_{\downarrow}^{\dagger} (\vect r) \phi_{\downarrow} (\vect r) \phi_{\uparrow}^{\dagger} (\vect r) \phi_{\uparrow} (\vect r)~. 
\end{align}
This model preserves \textit{time-reversal symmetry} and the $z$-component of the total angular momentum $L_z^T = \sum_i \ell_i$. Its phase diagram was obtained in Ref.~\onlinecite{dfOnSphere}. For strong repulsive interactions, the system is in a ferromagnetic phase with a two-fold degenerate ground state (weakly split in a finite system). We use the even combination with negative magnetization as state $B$ and massive Dirac fermions with $M_A=-2$ or $M_A=2$ for $A$. Our results for $12$ particles and $24$ single-particle states are shown in Fig.~\ref{fig: Many-body RelES cutoff3}. Despite the relatively small system size, the ES clearly identifies state $B$. If its magnetization matches the sign of the mass in $A$, there is a large gap in the ES [Fig.~\ref{fig: Many-body RelES cutoff3}(a)]. For opposite signs, it is gapless and exhibits a left-moving chiral mode [Fig.~\ref{fig: Many-body RelES cutoff3}(b)]. These spectra qualitatively agree with the analogous free-fermion spectra
[Fig.~\ref{fig: Many-body RelES cutoff3}(c,d)].

{\bf Relative Hamiltonians.} Having established the utility of relative ES numerically, we return to its analytical interpretation. It is illuminating to construct a parent Hamiltonian whose ground state is the relative wave function. We thus define the \textit{relative Hamiltonian} as the parent Hamiltonian $H^B$ of a surface phase $B$ expressed in terms of the excitations $c^{A}_{i,\pm}$. Since $A$ and $B$ are surface states by assumption, the relative Hamiltonian is bounded from below.\footnote{The Hamiltonians $H^{A,B}$ of any two surface states differ only within a finite energy window and the ground state of $H^A$ is the $c^{A}_{i,\pm}$ vacuum. }

As a concrete example, we consider massive Dirac fermions, described by Eq.~\eqref{eqn.massivedirac} for $A$. Using Eq.~\eqref{eqn.uandv}, we write
\begin{equation}
H_{M_A}=\sum\nolimits_{\vect k}\epsilon c_{\vect k,\pm}^{A,\dagger} c_{\vect k,\pm}^A,\quad 
 c^A_{\vect{k},\pm} =\frac{1}{2\sqrt{\epsilon_A}}
 \left[v_{\vect k}^*\phi_{\vect k}\pm u_{\vect k}^*\phi^\dagger_{-\vect k}\right]~,
\end{equation}
and the ground state satisfies $c^A_{\vect{k},\pm}|\Phi_{M_A}\rangle=0$. For $B$, we take Dirac fermions with a mass $M_B$, also described by Eq.~\eqref{eqn.massivedirac}. Expressing their Hamiltonian in terms of $c^A_{\vect k,\pm}$,
we find $H_{M_B} = H^+ + H^-$ with
\begin{equation}
 H^\pm
=\sum_{\vect k} E^\pm_k
 c_{\vect k,
\pm}^{A,\dagger}
 c^A_{\vect k,\pm}+\sum\nolimits_{\vect k}\left[\Delta^\pm_{\vect k}c^A_{\vect k,\pm}c_{-\vect k,\pm}^A+\text{H.c}\right]~\label{hrel}
\end{equation}
where $E^\pm_k=\epsilon_A+M_A(M_B-M_A)/\epsilon_A $ and $\Delta^\pm_{\vect k}= \pm (M_A-M_B) k e^{i\varphi_{\mathbf{k}}}/\epsilon_A $. 

For either choice of sign, Eq.~\eqref{hrel} describes a superconductor of spinless fermions with chiral $p$-wave pairing. The U(1) symmetry of the surface states $A$ and $B$ is reflected in $E^+=E^-$ and $|\Delta^+|=|\Delta^-|$. For an equal sign of the masses $M_A$ and $M_B$, the functions $E^\pm_k$ are positive for all $k$. The chemical potential of $c^A_{{\vect k},\pm}$ thus lies beyond the bottom of the band and $H_\pm$ each describes a topologically trivial strongly-paired superconductor. For opposite signs, the chemical potential lies within the band, and the superconductors are topological. 
A boundary where the mass changes sign hosts two chiral Majorana fermions or, equivalently, one chiral complex fermion.

As a second example, we take $A$ as before and choose $B$ as a superconducting surface state described by Eq.~\eqref{eqn.scprox}. We find $H^\text{SC}_{\Delta_s}=H^++ H^-$ with
$E^\pm_k = k^2/\epsilon_A \pm \Delta_s$ and $\Delta^\pm=k e^{\pm i \varphi_{\mathbf{k}}} \Delta_s$. For any non-zero $\Delta_s$, one of the flavors is strongly paired, and the second is weakly paired. Thus, there is always a single chiral Majorana at the interface of $H_{\Delta_s}$ with the reference vacuum. In Appendix~\ref{app: relative hamiltonian for sc sc}, we explicitly derive the relative Hamiltonian when both $A$ and $B$ both are paired states.

Table~\ref{table: relative hamiltonians} summarizes the possible relative Hamiltonians for various choices of $A$,$B$. We note that a simple analytical expression for the ES of $H^+$ on an infinite cylinder was obtained in Ref.~\onlinecite{ChiralSuperfluidsReadDubail} for a particular choice of the functions $E^+_k,\Delta^\pm_{\vect k}$.\cite{Special_Point} For generic parameters and different geometries, a similar expression is unavailable. Still, for free-fermion states, the ES can be efficiently obtained from correlation matrices.

\begin{table}
\centering

\caption{The relative Hamiltonians realize topological $p
\pm ip$ superconductors or topologically-trivial strongly paired (SP) superconductors depending on the type and sign of the mass terms. The helical edge state arising for opposite superconducting mass terms is protected by time-reversal symmetry.}\

\begin{tabular} {l|l|l}
\hline
\hline

& Magnetic gap $M^A>0$ & Pairing gap $\Delta^A>0$\\
\hline
\hline
$M^B>0$ & SP and SP
& $p-ip$ and SP\\
\hline
$M^B<0$ & $p+ip$ and $p+ip$
& $p+ip$ and SP\\
\hline
$\Delta^B<0 $&$p+ip$ and SP & $p+ip$ and $p-ip$\\
\hline
$\Delta^B>0 $&$p-ip$ and SP & SP and SP\\
\hline\hline
\end{tabular}

\label{table: relative hamiltonians}

\end{table}

{\bf Discussion and conclusions.}
We have generalized entanglement spectroscopy as a tool for identifying phases of matter to SPT surface states. Mapping the boundary between two surface states onto the analogous edge of a non-anomalous state allowed us to invoke the standard ES-edge correspondence. We have demonstrated the utility of relative ES via large-scale numerical simulations of free-fermion and interacting systems. 

The most immediate applications of our analysis arise in numerical studies of SPT surfaces, where it may identify non-trivial phases. For example, symmetry preserving surface topological orders are known to be possible on 3D TI and 3D topological superconductor surfaces \cite{StrongInteractions1, StrongInteractions2, StrongInteractions3, StrongInteractions4, StrongInteractions5,3dscsurfaceLukasz,3dscsurfaceChong,3dscsurfaceMax,3dscsurfaceJeffrey}. Relative ES and the corresponding \textit{relative entanglement entropy} would be the natural tools for their numerical identification in candidate systems.

Anomalies also arise in systems other than SPT surfaces. For example, in fractional quantum Hall systems at half-filling, an anomalous particle-hole symmetry plays an analogous role to time-reversal on the 3D TI surface.\cite{sonDirac2015} Specifically, the \textit{orbital} ES is mirrored under a particle-hole transformation. As such, it must be non-chiral for a particle-hole symmetric state and cannot represent its edge with a trivial vacuum. We expect that a variant of the relative ES may be advantageous in the context of such states. 

Additionally, the methods developed here can prove beneficial for analyzing various one-dimensional spin systems. It is often convenient to fermionize these models via a Jordan-Wigner transformation. The non-locality of this mapping results in an anomalous theory of fermions\cite{dualitywebz2}, suggesting that relative ES will be the appropriate tool.

Finally, ES may also be valuable for non-anomalous systems. We note that for a band insulator, the relative ES between states with $n-1$ and $n$ filled band encodes exactly the contribution of the $n$th band. More generally, relative ES may help identify interacting states with complex edge structures that are not readily deduced from their (many-body) ES. In fact, a variant of this is already known in quantum Hall systems. There, performing a particle-hole transformation prior to a real-space cut\cite{REESChandran,REESDubail,REESSimon} amounts to computing the relative ES with a $\nu=1$ quantum Hall state. As a consequence, the ES of `hole-conjugate states' such as the anti-Pfaffian greatly simplify. More generally, obtaining the boundary of an unknown state with multiple two-dimensional phases may help disentangle the contributions from multiple edge modes. 

{\bf Acknowledgements.} We are indebted to Misha Yutushui for sharing his code for exact diagonalization on the 3D TI surface, and to Jason Alicea and Lesik Motrunich for illuminating discussions on this topic. This work was partially supported by the Deutsche Forschungsgemeinschaft (CRC/Transregio 183)
and the Israel Science Foundation (2572/21).

\bibliographystyle{apsrev4-2}
\bibliography{bibtex}

\newpage

\appendix

\section{Surface states of spherical topological insulators}
\label{app: surface states of spherical 3DTI spectrum}

\begin{figure}[h]
 \includegraphics[width=0.8\columnwidth]{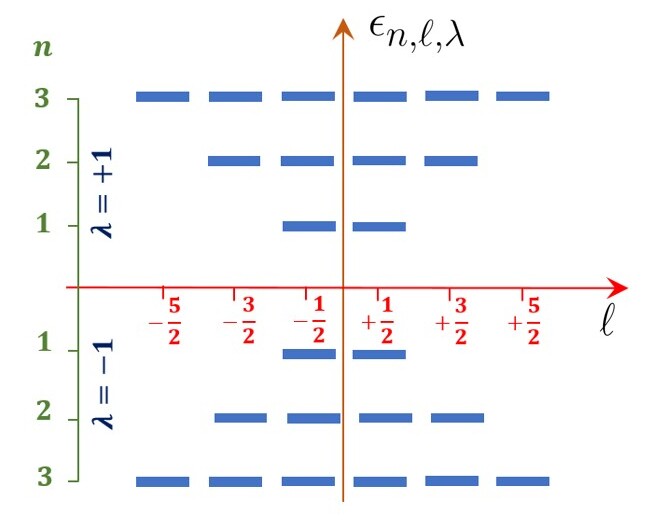}
 \caption{Spectrum of a spherical $3$D topological insulator surface. States are labeled by the positive integer $n$, the angular momentum $\ell \in[-n+1/2,n-1/2]$ and the particle-hole index $\lambda=\pm$. A cutoff in the $n$ quantum number preserves angular momentum.}
 \label{fig: spectra of dirac fermions}
\end{figure}

The single-particle eigenstates on a spherical TI surface \cite{dfOnSphere, LLQuantizationSphere} are given by
\begin{equation}
 \phi_{n,\ell,\lambda} = \frac{1}{\sqrt{2}}
 \begin{pmatrix}
 Y_{1/2,n+1/2,\ell} \\
 \lambda Y_{-1/2,n+1/2,\ell}~
 \end{pmatrix},
\end{equation}
where $Y$ are the monopole harmonics~\cite{JainBook, LLonSphere}. The corresponding energies are $\epsilon_{n,\ell,\lambda} = \lambda n$, where $\lambda = \pm 1$ and $n$ is a positive integer. The angular momentum $\ell$ takes half-odd integer values in the interval $[-n+1/2,n-1/2]$. For any numerical calculation, we implement a cutoff by retaining only $n\leq \Lambda$. The spectrum for $\Lambda=3$ is shown in Fig.~\ref{fig: spectra of dirac fermions}.

\section{Energy spectra of interfaces on spherical topological insulator surfaces}
\label{app: physical energy spectrum of spherical 3DTI}

\begin{figure}
 \centering
 \includegraphics[width=\columnwidth]{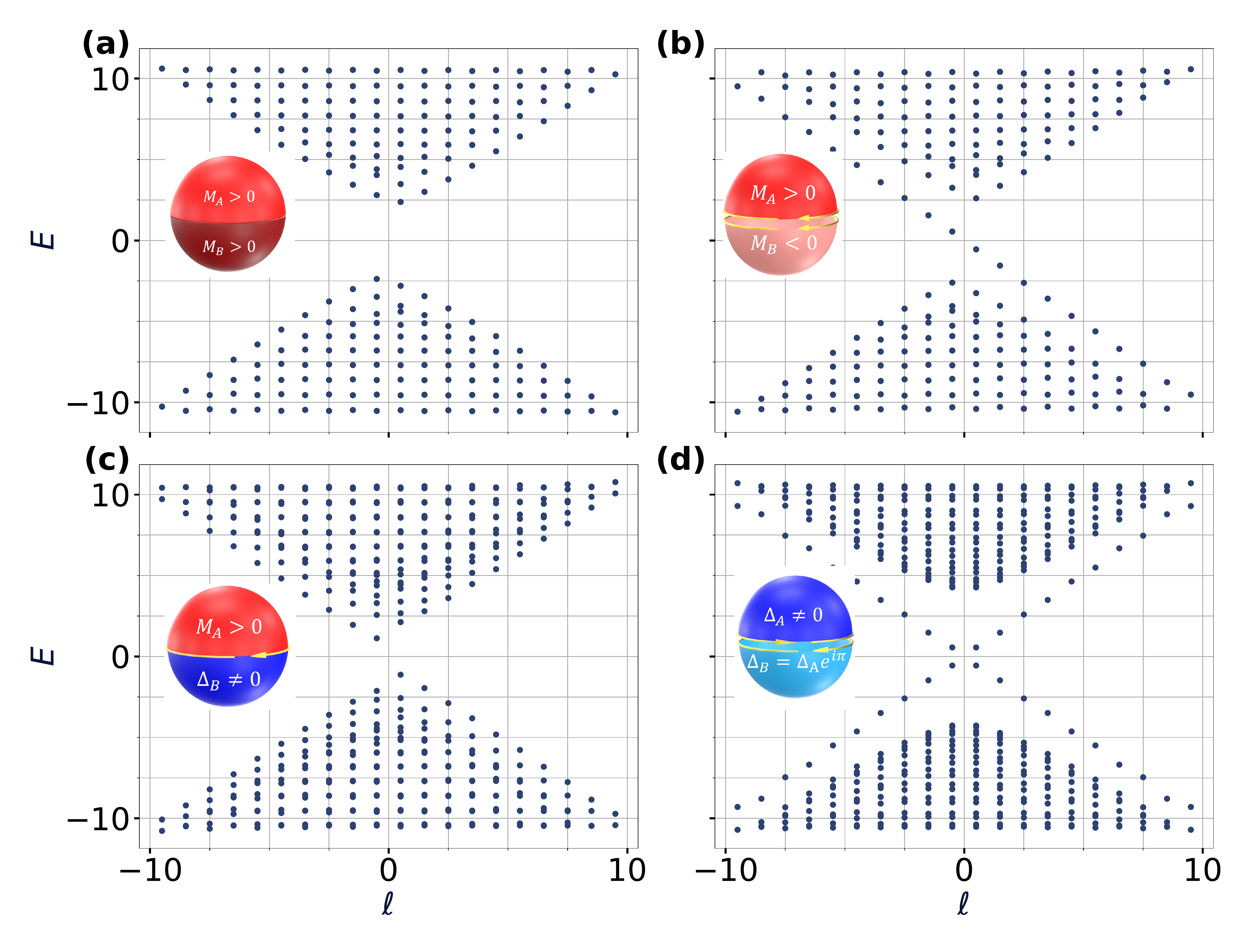}
 \caption{ Physical edge spectra of spherical $3$D topological insulator surfaces. In panels (a) and (b), both hemispheres are magnetically gapped. When the masses have equal signs, the spectrum is gapped (a). When they have opposite signs, it hosts a gapless chiral complex fermionic edge mode (b). In panel (c), the lower hemisphere is a proximity-induced $s$-wave superconductor, hence the spectrum hosts a chiral Majorana mode. In panel (d), both the systems are $s$-wave superconductors with a phase difference of $\pi$. The spectrum hosts two counter-propagating gapless Majorana channels.}
 \label{fig: DF physical Edge Spectrum}
\end{figure}

The interface of two distinct surface phases hosts gapless edge channels. These edge states carry signatures of the surface phase. Fig. \ref{fig: DF physical Edge Spectrum} shows energy spectra for various surface states on the upper and lower hemisphere. Panel (a) shows the spectrum when both hemispheres realize TRS-broken phases with a positive magnetic mass. The resulting spectrum is gapped since both hemispheres are in the same topological state. In panel (b), the upper hemisphere has 
a positive mass, and the lower hemisphere has a negative mass. The resulting spectrum includes a chiral mode that corresponds to a complex fermion. In panel (c), the upper hemisphere is as before, while the lower forms a time-reversal invariant superconductor. The resulting spectrum contains a chiral Majorana mode. In panel (d), both hemispheres are $s$-wave superconductors, with a phase difference of $\pi$. Here, the spectrum describes helical Majorana edge states.

\section{Cutoff dependence of entanglement spectra for anomalous surface states}
\label{app: cutoff dependence of es and reles}
\begin{figure}[h]
 \centering
 \includegraphics[width=\columnwidth]{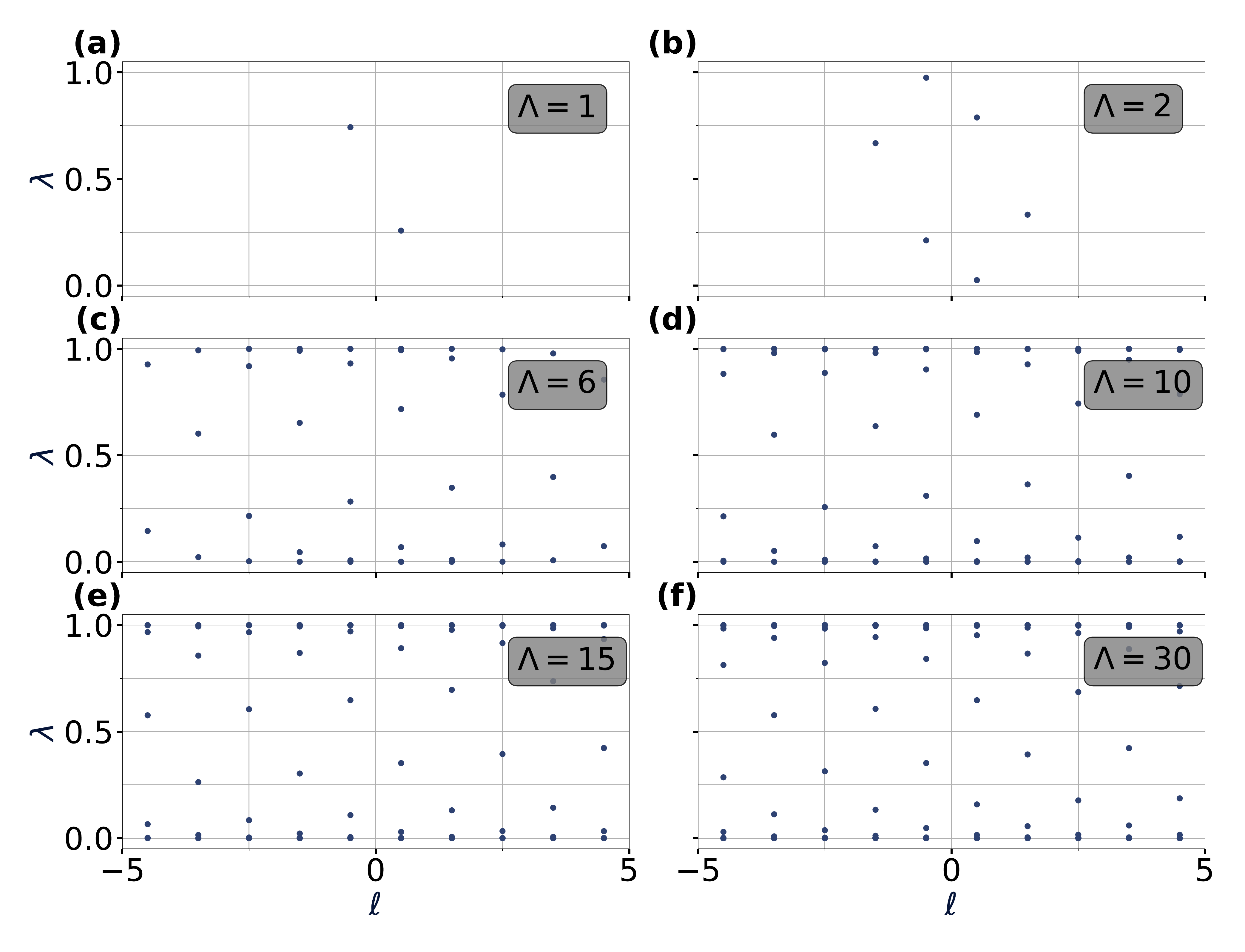}
 \caption{The entanglement spectra of anomalous surface states are strongly sensitive to the short-distance cutoff. Panels (a)-(f) show the entanglement spectra for $3$D topological insulator surfaces with a magnetic mass $M=2$ and cutoff values $\Lambda=1,2,6,10,15,30$, respectively}
 \label{fig: ES Cutoff Dependence}
\end{figure}
\begin{figure}[h]
 \centering
 \includegraphics[width=\columnwidth]{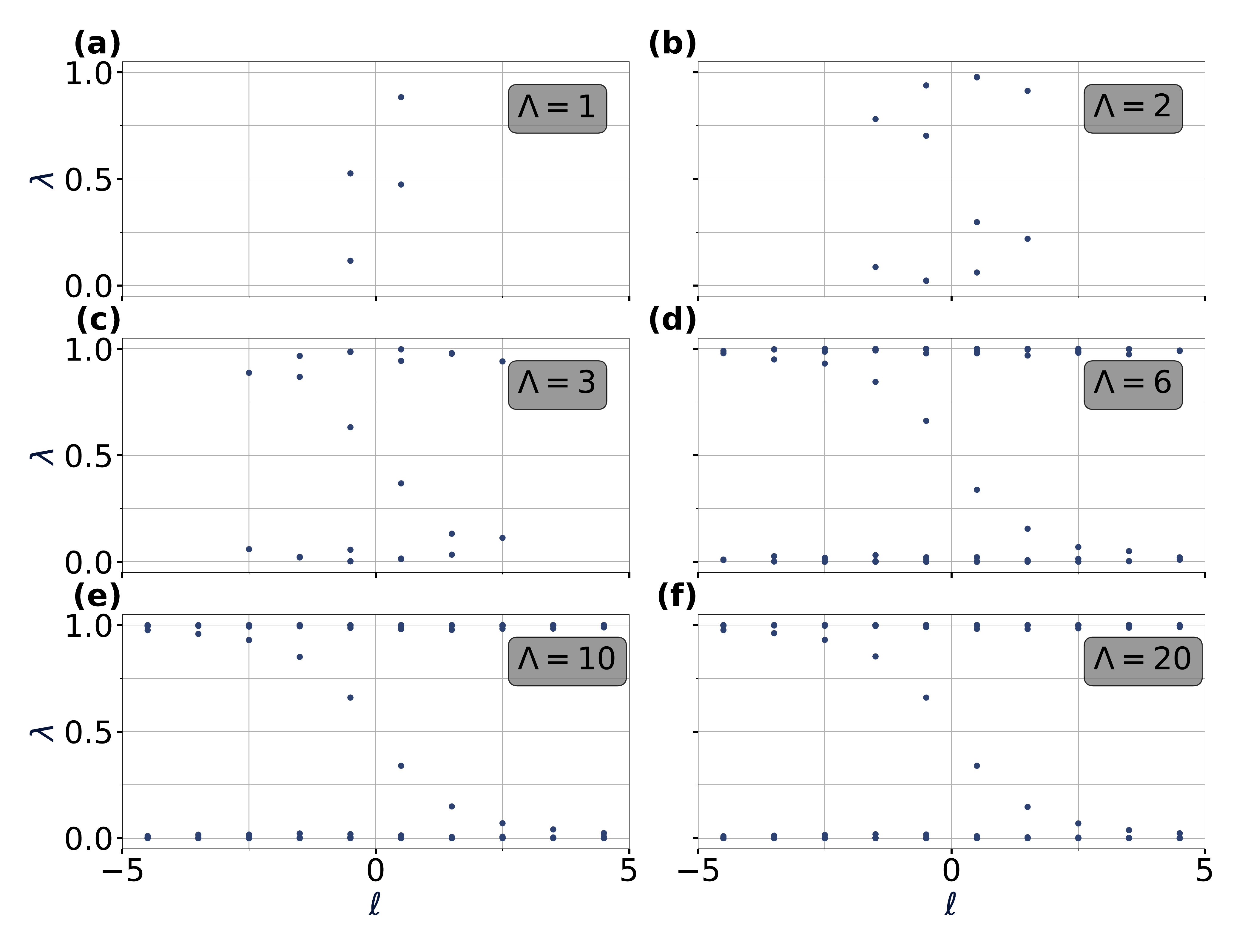}
 \caption{The relative entanglement spectra of anomalous surface states depend only weakly on the cutoff. Panels (a)-(f) show the spectra for $M_A=2$, $M_B=-2$ and $\Lambda=1,2,3,6,10,20$, respectively. The low-lying states become insensitive to the cutoff for $\Lambda \gtrsim 3$. }
 \label{fig: RelES cutoff Dependence}
\end{figure}

The cutoff dependence of the ES for a magnetically gapped TI surface is shown in Fig.~\ref{fig: ES Cutoff Dependence}. The number of low-lying entanglement-energy levels ($\lambda \approx 0.5$) increases linearly with the cutoff. As such, the ES is dominated by non-universal features.
By contrast, the low-lying states in the relative ES quickly converge to their $\Lambda \rightarrow \infty$ values, see Fig.~\ref{fig: RelES cutoff Dependence}. There is an appreciable entanglement gap already for a cutoff as small as $\Lambda=3$.

\section{Entanglement spectra of continuum systems from their correlation matrix}
\label{app: zeroes of the decomposed hamiltonian}

The ES of a free fermion system is fully encoded in the correlation matrix.\cite{correlationMatrix, correlationMatrix2, correlationMatrix3, correlationMatrix4, correlationMatrix5, correlationMatrixBdG, correlationMatrixBdG2, correlationMatrixBdG3} The latter can be efficiently obtained by inverting the single-particle Hamiltonian $H$. In continuum models, such as the one describing the 3D TI surface, there is a minor subtlety that is absent on lattice systems. To illustrate this issue, let $\psi_i(\vect x)$ with $i=1 \ldots N$ be an orthonormal basis of single particle states for the full system. The projections of $\psi_i$ onto any subsystem are generically not orthonormal. Still, standard methods readily obtain orthonormal functions $\psi^A_i$ in $A$ and $\psi^B_i$ in $B$ for the projected $\psi_i$. As a result, we obtain the decomposition
\begin{equation}
 \psi_i(\vect r) = \sum_j \left[\alpha_{ij} \psi_j^A(\vect r) + \beta_{ij} \psi_j^B(\vect r)\right].\label{app.decomp}
\end{equation}

Notice that the $N$ states of the original system are encoded in $2N$ states after this decomposition. Consequently, Eq.~\eqref{app.decomp} is not invertible, and the correlation matrix for states within $A$, cannot be directly obtained from the full Hamiltonian. Equivalently, inserting Eq.~\eqref{app.decomp} into any Hamiltonian enlarges the Hilbert space by $N$ single-particle states with zero eigenvalue. 

To circumvent this problem, we enlarge the original Hilbert space by $N$ single-particle states at infinite energy. Specifically, let $v_i = (\alpha^0_{i,1}\ldots \alpha^0_{i,N}\ \beta^0_{i,1}\ldots \beta^0_{i,N})$ be the null-vectors of the matrix 
\begin{equation}
 \mathcal{M}
 =
 \begin{pmatrix}
 \alpha_{11} & \hdots & \alpha_{1N} & \beta_{11} & \hdots & \beta_{1N} \\
 \vdots & \vdots & \vdots & \vdots & \vdots & \vdots \\
 \alpha_{N1} & \hdots & \alpha_{NN} & \beta_{N1} & \hdots & \beta_{NN}
 \end{pmatrix}~.
\end{equation}
Then the wave functions

\begin{equation}
 \psi^0_i(\vect r) \equiv \sum_j \left[\alpha_{ij}^0 \psi_j^A(\vect r) + \beta_{ij}^0 \psi_j^B(\vect r)\right].\label{app.decomp2}
\end{equation}
supplement Eq.~\eqref{app.decomp} to yield an invertible transformation between $\psi_i,\psi_i^0$ and $a_i,b_i$. Finally we modify the single particle Hamiltonian according to $H \rightarrow H + H'$ with $\langle \psi_i| H' |\psi_j \rangle =\langle \psi_i^0| H' |\psi_j \rangle =0$ and $\langle \psi_i^0| H' |\psi_j^0 \rangle =m \delta_{ij}$ with $m\rightarrow \infty $ to ensure that the additional states do not affect the spectrum or the correlations. After these modifications, the correlation matrix and ES can be computed as in lattice systems.

\section{Evolution of relative entanglement spectra with magnetic mass}
\label{app: evolution of reles}
\begin{figure}[ht]
 \centering
 \includegraphics[width=\columnwidth]{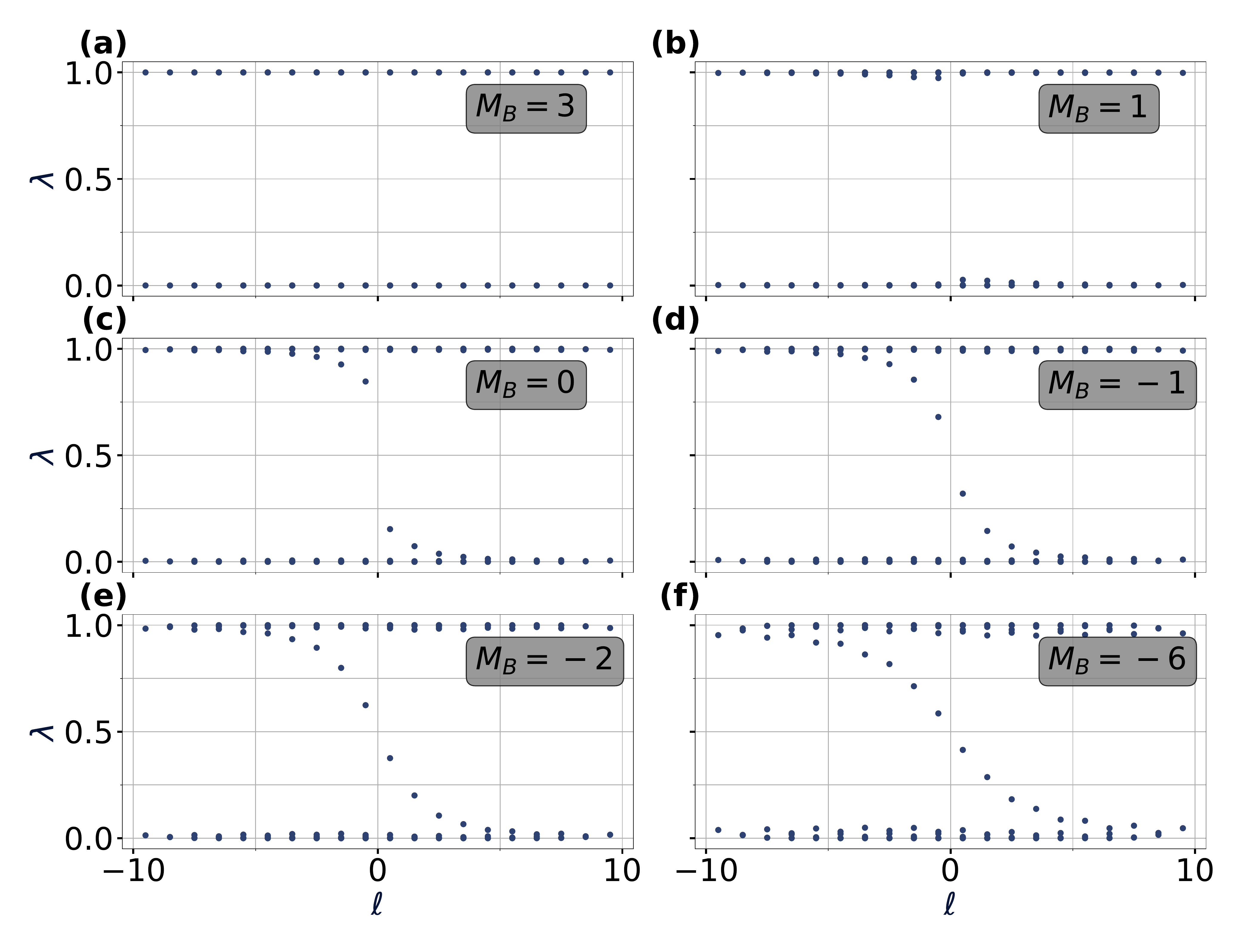}
 \caption{Evolution of relative entanglement spectra between $M_A = 3$ and different choices of $M_B$. For opposite signs of the masses, a gapless edge crosses the entanglement gap.} 
 \label{fig: reles evolution}
\end{figure}
We show the relative ES for $M_A=3$ as a function of $M_B$ in
Fig.~\ref{fig: reles evolution}. For equal signs of $M_A$ and $M_B$, the gap in the relative ES begins to close as $M_B \rightarrow 0$. For opposite signs, the edge state emerges and becomes better resolved with increasing $M_B$.

\section{Relative Hamiltonian for two superconducting surface states}
\label{app: relative hamiltonian for sc sc}

We derived the relative Hamiltonians for a magnetically-gapped reference system in the main text. Alternatively, a superconducting surface can also serve as a reference system. To identify its excitations, we diagonalize the model of Eq.~\eqref{eqn.scprox}, i.e., the Hamiltonian
\begin{equation}
 \begin{split}
 H^\text{SC}_{\Delta_s} 
 & = \sum_{\mf{k}} \phi_{\mf{k},\uparrow}^{\dagger} (k e^{-i\varphi_{\mf{k}}} \phi_{\mf{k},\downarrow} + \Delta_s \phi_{-\mf{k},\downarrow}^{\dagger}) \\
 & \quad \qquad + (k e^{i\varphi_{\mf{k}}} \phi_{\mf{k},\downarrow}^{\dagger} + \Delta_s^{*} \phi_{-\mf{k},\downarrow}) \phi_{\mf{k},\uparrow}~.
 \end{split}
 \label{eq: pairing hamiltonian with deltas}
\end{equation}
We begin with a Bogoliubov transformation to introduce new fermions $\zeta_{\mf{k},\downarrow}$
\begin{equation}
 \zeta_{\mf{k},\downarrow} = \frac{1}{\epsilon_\Delta} (k e^{-i\varphi_{\mf{k}}} \phi_{\mf{k},\downarrow} + \Delta_s \phi_{-\mf{k},\downarrow}^{\dagger}),
 \label{eq: new fermion}
\end{equation}
with $\epsilon_\Delta = \sqrt{k^2 + |\Delta_s|^2} $. They satisfy the usual anticommutation relations

\begin{equation}
 \begin{split}
 & \{ \zeta_{\mf{k},\downarrow}^{\dagger}, \zeta_{\mf{k'},\downarrow} \} = \delta_{\mf{k},\mf{k}'}~, \\
 &\{ {\zeta_{\mf{k},\downarrow},\zeta_{\mf{k'},\downarrow}} \} = \{ \zeta_{\mf{k},\downarrow}^{\dagger}, \zeta_{\mf{k}',\downarrow}^{\dagger} \} = 0~, \\ 
 & \{ \phi_{\mf{k},\uparrow}^{\dagger}, \zeta_{\mf{k'},\downarrow} \} = \{ \phi_{\mf{k},\uparrow}^{\dagger}, \zeta_{\mf{k'},\downarrow}^{\dagger} \} = 0~. 
 \end{split}
\end{equation}
Substituting Eq.~\eqref{eq: new fermion} into Eq.~\eqref{eq: pairing hamiltonian with deltas} yields
\begin{equation}
 H_{\Delta_s}^{\text{SC}} = \sum_{\mf{k}} \epsilon_\Delta (\phi_{\mf{k},\uparrow}^{\dagger} \zeta_{\mf{k},\downarrow} + \zeta_{\mf{k},\downarrow}^{\dagger} \phi_{\mf{k},\uparrow}).
\end{equation}
Introducing another set of fermion operators as \begin{equation}
 \chi_{\mf{k},+} = \frac{1}{\sqrt{2}} (\phi_{\mf{k},\uparrow} + \zeta_{\mf{k},\downarrow})~, \quad \chi_{\mf{k},-} = \frac{1}{\sqrt{2}} (\phi_{-\mf{k},\uparrow}^{\dagger} - \zeta_{-\mf{k},\downarrow}^{\dagger})~,
 \label{eq: diagonal opertors of pairing pairing reference hamiltonian}
\end{equation}
we finally obtain
\begin{align}
 H_{\Delta_s}^{\text{SC}} = \sum_{\mf{k}} \epsilon_\Delta (\chi_{\mf{k},+}^\dagger\chi_{\mf{k},+}+\chi^\dagger_{\mf{k},-}\chi_{\mf{k},-})~.
 \end{align}

Consider now a surface state with different pairing $\Delta'_s$. In terms of the operators defined by Eq.~\eqref{eq: diagonal opertors of pairing pairing reference hamiltonian}, its Hamiltonian is
\begin{align}
 & H_{\Delta'_s}^\text{SC} = H^{\pm}
 + \sum_{\mf{k}} (\Delta_{\mf{k}}^M \chi_{\mf{k},+}^{\dagger} \chi_{-\mf{k},-}^{\dagger} + \text{H.c.})~,
\end{align}
where $H^\pm$ is given by Eq.~\eqref{hrel} with dispersion
$E_{k}^{\pm} = (k^2 + \frac{\Delta'_s \Delta_s^{*} + \Delta'_s{}^{*} \Delta_s}{2})/\epsilon_\Delta$, and mean-field pairing $\Delta_{\vect{k}}^{-} = k e^{ i\varphi_{\vect{k}}} (\Delta_s - \Delta'_s)/2\epsilon_\Delta$, $\Delta_{\vect{k}}^{+} = (\Delta_{\vect{k}}^{-})^{*}$. The final term, $\Delta_{\vect{k}}^{M} = (\Delta'_s{}^{*} \Delta_s - \Delta'_s \Delta_s^{*})/\epsilon_\Delta$, vanishes when the phases of $\Delta_s'$ and $\Delta_s$ differ by $\pi$. In that case, $H^+$ and $H^-$ decouple. Each describes a superconductor with $k_x \pm i k_y$ pairing and negative `chemical potential' $(\Delta'_s \Delta_s^{*} + \Delta'_s{}^{*} \Delta_s)/2$. The boundary of this system hosts a pair of counter-propagating Majorana modes. For other phase differences, the two Majorana modes gap out.

\section{Relative entanglement spectra for quantum spin Hall boundary states}
\label{app: ring geometry 2D qsh}

\begin{figure}[!h]
 \includegraphics[width=\columnwidth]{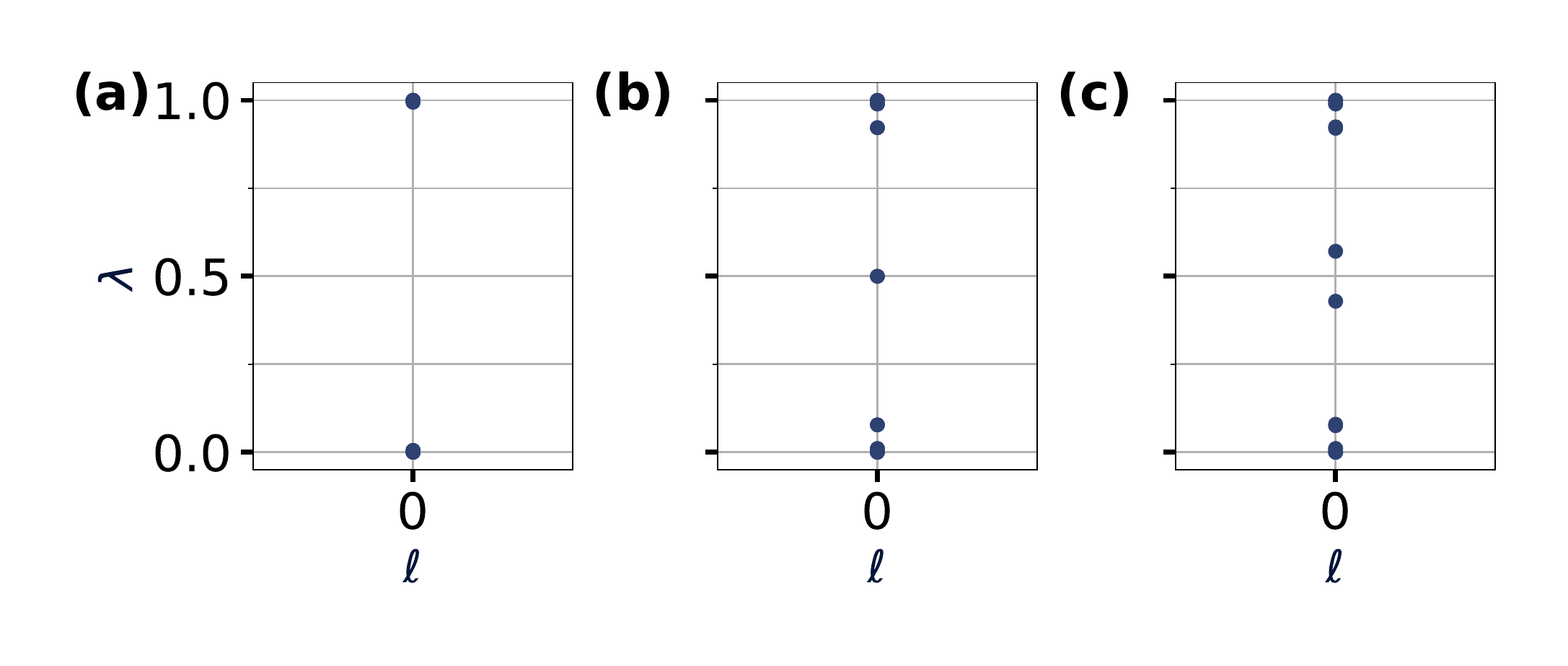}
 \caption{A 2D quantum spin Hall system on a disk hosts a gapless Dirac fermion on its boundary, i.e., a one-dimensional ring. Breaking time-reversal symmetry opens a gap, as for the 3D topological insulator. The relative entanglement spectrum for a massive edge with the equal sign is gapped, as shown in (a) for $M_A=2$ and $M_B=4$. By contrast, there is a zero mode (state with $\lambda=0.5$) for opposite masses, as shown in panel (b) for $M_A=2$ and $M_B=-4$. This zero-mode splits when a small pairing term is added to the probing system of panel (c).}
 \label{fig: reles qsh} 
\end{figure}

The relative ES applies to any system that permits a gapped free-fermion reference state. As an additional example, we provide results for the two-dimensional quantum spin Hall (QSH) phase on a circular disk. Its physical boundary hosts a single one-dimensional Dirac fermion. In Fig.~\ref{fig: reles qsh}, we show the zero-dimensional relative ES when this boundary becomes gapped by magnetic masses $M_A$,$M_B$. The relative ES is gapped for equal signs of the two masses and exhibits a state at zero entanglement energy ($\lambda=0.5$) for opposite signs. Adding a small pairing term splits this zero mode, see Fig.~\ref{fig: reles qsh}(c). These properties correctly reproduce the known behavior of QSH edge states, where the interface between opposite masses hosts a complex fermion zero mode.

\end{document}